\documentstyle[12pt,epsfig,axodraw,amssymb]{article}

\newcommand{\sm}[1]{{\scriptscriptstyle #1}}

\newcommand{\be}{\begin{equation}}
\newcommand{\ee}{\end{equation}}
\newcommand{\ba}{\begin{eqnarray}}
\newcommand{\ea}{\end{eqnarray}}
\begin{document}

\title{Galactic anisotropy of Ultra-High Energy Cosmic Rays 
produced by CDM-related mechanisms} 
\author{  S.~L.~Dubovsky\footnote{E-mail: sergd@ms2.inr.ac.ru},
  P.~G.~Tinyakov\footnote{E-mail: peter@ms2.inr.ac.ru}\\
  {\small\it
     Institute for Nuclear Research of
         the Russian Academy of Sciences, 117312 Moscow, 
Russia
  }}
\date{}
\maketitle

\begin{abstract}
We briefly review current theoretical and experimental status of
Ultra-High Energy Cosmic Rays. We show that ``top-down'' mechanisms of
UHE CR which involve heavy relic particle-like objects predict
Galactic anisotropy of highest energy cosmic rays at the level of
minimum $\sim 20\%$. This anisotropy is large enough to be either
observed or ruled out in the next generation of experiments.
\end{abstract}

This talk is devoted to a specific signature of some mechanisms
of production of Ultra-High Energy Cosmic Rays~(UHE CR)~\cite{mi},
related to non-central position of the Earth in our Galaxy. We start
by brief review of the current theoretical and experimental status of
UHE CR. We then describe the class of mechanisms to which
our arguments apply and
explain the origin of the anisotropy of the UHE CR produced by these
mechanisms. Finally, we estimate the anisotropy and give our
conclusions. 

UHE CR consist of the most energetic particles available for
physicists at present, with energies in excess of $10^{19}$ eV. At
these energies the flux of cosmic rays which falls like $E^{-3}$ is
very small, roughly 1 event per km$^2$ per century.  The number of
events observed so far is well below hundred. The primary particle
content is currently unknown. 

The special interest in cosmic rays with energies higher than
$10^{19}$ eV is related to the cutoff predicted in the spectrum in
this energy range, the so-called GZK cutoff~\cite{GZK}. The origin of
this effect is easy to understand assuming that primary particles are
protons. At energies exceeding $E_{GZK}\sim 4\times 10^{19}$ eV a
proton propagating through cosmic microwave background rapidly
loses its energy due to resonant pion photoproduction. The proton
mean free path $R_{GZK} \sim 50$ Mpc is two orders of magnitude
smaller than the size of the visible part of the
Universe. Consequently, the flux of cosmic rays (assuming they are
protons) is expected to drop by two orders of magnitude at $E\sim
E_{GZK}$. Similar cutoff is expected for photons \cite{photonGZK}.

There are several experiments capable of detecting the low flux of UHE
CR, the largest being Akeno Giant Air Shower Array~(AGASA), Fly's Eye
I,II Experiments and Yakutsk Experiment. The common idea of these
experiments is observation of showers which
are produced in the collisions of primary UHE particles with
atmosphere. Detecting such a shower and reconstructing the energy and
arrival direction of a primary particle requires a large number of
detectors on the total area of order several decades of square
kilometers. AGASA, for example, consists of 111 scintillators,
 located on the area of 100~km$^2$.

The sensitivity of the experiments has just reached the energy range
where the GZK cutoff is expected. About 20 events with energies higher
than $E_{GZK}$ were observed. There were 8 events detected with energy
above $10^{20}$~eV. Two most reliable ones have the following
energies: $2.1^{+0.5}_{-0.4}\times 10^{20}$~eV (AGASA,~\cite{AGASA})
and $3.2^{+0.92}_{-0.94}\times 10^{20}$~eV (Fly's Eye,~\cite{Fly}).
While the typical angular resolution in these experiments is rather
high, of order 3$^{\circ}$, the energy resolution is only about 30\%.
As a result of poor statistics and low energy resolution, current data
are not enough to draw definite conclusion about the very existence of
the GZK cutoff. They, however, indicate that the expected sharp cutoff
is absent.

If confirmed, the absence of the cutoff would definitely be a hint for
a new physics. There may be two possible explanations: either primary
particles are some new particles (e.g. UHEcrons, \cite{farrar}) which do
not interact with cosmic microwave background, or the sources of UHE
CR are relatively close to us (i.e., within $\sim 50$~Mpc). In what
follows we concentrate on the latter possibility. On this way, the
main problem is a {\em mechanism} of production of cosmic rays with
such a huge energy.

Regardless of the location of the sources, possible mechanisms of
production of UHE CR are naturally divided into two classes --- the
astrophysical and particle-physics ones. The astrophysical mechanisms
typically employ acceleration of charged particles in strong magnetic fields.
There are severe constraints on maximum energy to which a particle can
be accelerated at a given value of magnetic field and the size of
astrophysical object \cite{hillas},
$$
BL>{E\over 10^{15} \mbox{eV}}{1\over Z\beta}\;,
$$
where $B$ is the magnetic field in $\mu$G, $L$ is the size in
parsecs, $Z$ is the charge of the particle and $\beta$ is the speed of
the shock wave. There are very few astrophysical objects (for
instance, active galactic nuclei~\cite{AGN,hot} or hot spots of radio
galaxies~\cite{hot}) which are believed to be able to accelerate
particles to energies of order $10^{20}$ eV.  If such an object
located within 50 Mpc from us were the source of UHE CR it would be
identified. Thus, in order to reconcile the astrophysical mechanism with
the absence of the GZK cutoff one would have to consider 'exotic'
primary particles.

The particle-physics mechanisms typically involve decays of heavy
par\-ticle-like objects, ``X-particles'', either primordial or recently
produced in the process of the evolution of cosmological defects. The
mechanisms of the former type we call ``CDM-related''. Their
characteristic feature is that the sources of UHE CR are distributed
in the Universe in the same way as Cold Dark Matter (CDM), i.e., they are
concentrated in galactic halos as a result of gravitational clustering
at the stage of galaxy formation\footnote{Note that distribution of
X-particles inside a halo does not need to follow that of CDM as it is
determined by the interactions of particles with each other and with
matter.} \cite{clustering}. 
Numerically it means that the average densities of
sources of UHE CR in the Universe
$\bar{n}$ and in the Galactic halo $\bar{n}_h$ 
are related in the same way as the average densities of the matter in
the Universe and in the Galaxy,
\begin{equation}
{\bar n \over \bar n_h} \simeq {\Omega_{\sm{CDM}} \rho_{\sm{crit}} \over
\bar\rho_{\sm{halo}} } \sim 10^{-5}.
\label{n/n_h}
\end{equation}
According to this definition, any
mechanism involving primordial massive particles which are
non-relativistic at the time of galaxy formation, is CDM-related.  On
the contrary, mechanisms where X-particles are permanently produced by
topological defects like cosmic strings or 'cosmic necklaces' (for a
review see Refs.~\cite{topolog}) are not CDM-related.

For the sake of completeness consider briefly some of the CDM-related
mechanisms discussed in the literature. Simplest ones are based on
decays of heavy long-living
particles~\cite{ellis,X-KR,X-Ber}. Regardless of their nature, the
mass and lifetime of these particles must lie in a certain range.
Indeed, the flux $F$ of UHE CR resulting from decays of the relic
X-particles is
\begin{equation}
\label{flux}
{dF\over d\log{E}}\sim {n_X\over \tau_X}R_{GZK}N\;,
\end{equation}
where $n_X$ is the average number density of X-particles and $\tau_X$
is their lifetime. $N$ is average multiplicity of UHE CR produced in
one decay; it equals to the number of produced jets times the
fragmentation function. The expected value of $N$ lies in the range
$N\sim 10\div1000$.  Thus, the mass of X-particles should satisfy
\[
m_X\gtrsim 10^{13} \mbox{GeV}.
\]
Such heavy particles can be produced either during reheating (if the
reheating temperature was of order $m_X$) or directly from vacuum
fluctuations during inflation~\cite{X-KT,X-KolbRiotto}. 

The lifetime of X-particles can be bounded from the requirements that 
they produce the observed flux of UHE CR and do not overclose the
Universe. One gets \cite{X-KR} 
\[
10^{10} \mbox{yr} \lesssim \tau_X \lesssim
10^{22}\mbox{yr} .
\]
It is difficult to explain naturally such a long but finite lifetime.
One may speculate that decay of X-particles, which are otherwise
stable, is due to instanton-type~\cite{X-KR} or wormhole~\cite{X-Ber}
effects. Detailed discussion of the hypothesis that the UHE cosmic rays
result from decays of metastable massive relic dark matter particles
halo can be found, e.g., in ref.~\cite{sabib}.

Another potential mechanism of UHE CR production is
monopole-anti\-mo\-no\-pole annihilation~\cite{monopoles}. If
monopoles exist in Nature, one may expect that some of them are in the
form of monopole-antimonopole bound state (monopolonium). Certainly,
the ground state of monopolonium is very unstable. For example, in the
non-relativistic model the first Bohr radius of the monopolonium is
much less than the size of the monopole and monopole-antimonopole pair
should immediately annihilate. However, highly excited states of
monopolonium can be rather long-living. Estimates of~\cite{monopoles}
give $\tau_m\sim 40$~days for monopolonium of the size $r_m\sim 1$~fm
and $\tau_m\sim 10^{11}$~yr for $r_m\sim 1$~nm.

The scenario of UHE CR production by monopolonium is the
following. Monopolonium forms in highly excited state with $r_m\sim
1$~nm. Then it radiates light vector bosons and comes down to the
ground state after a time comparable to the age of the
Universe. Finally monopole and antimonopole annihilate and produce
heavy gauge bosons. Primary UHE particles appear as products of decays
of these bosons. Estimates of~\cite{monopoles} show that required the
abundance of the monopolonium can in principle be consistent with
experimental limits.

Now let us turn to the main topic of this talk, the specific signature
of the CDM-related mechanisms. We argue that, regardless of their
nature, all CDM-related mechanisms predict anisotropic flux of UHE CR
with the excess of at least 20\% towards the center of our
Galaxy~\cite{mi}.

The observed flux of UHE CR can be divided into Galactic and
extragalactic parts, 
\[
j = j_{\scriptsize\rm ext} + j_{h},
\]
where
\begin{equation}
j_h = C \int_{\scriptsize\rm halo} {d^3x \over x^2} n(x)
\label{j-halo}
\end{equation}
is the contribution of our Galaxy and 
\[
j_{\scriptsize\rm ext} = C \, 4\pi R_{\scriptsize\rm ext} \bar n
\]
has extragalactic origin. Here $R_{ext}=R_{\scriptsize\rm Universe}
\sim 4$~Gpc for energies below $E_{\scriptsize\rm GZK}$ and
$R_{\scriptsize\rm ext}\sim 50$~Mpc for energies above
$E_{\scriptsize\rm GZK}$. Note that the constant $C$ is the same in
both equations. 

\begin{figure}[htb]
\begin{center}
\epsfig{file=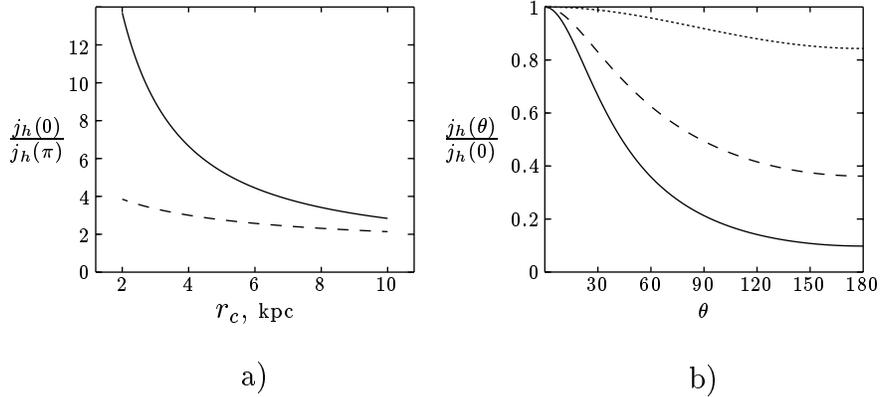,%
bbllx=100pt,bblly=535pt,%
bburx=257pt,bbury=695pt,
clip=}
\epsfig{file=fig1.ps,%
bbllx=285pt,bblly=535pt,%
bburx=460pt,bbury=695pt,
clip=}
\end{center}
\caption{a)~The anisotropy $j_h(0)/j_h(\pi)$ as a function of the core
size $r_c$ for the density profiles (\protect\ref{n=1/r^2}) (solid
line) and (\protect\ref{n=1/r}) (dashed line). b)~The corresponding
angular distributions at $r_c=5$~kpc. The dotted line shows the
angular distribution for $n(x)=\mbox{const}$ (i.e., when the
anisotropy is minimum).}
\end{figure}
The Galactic part of the total UHE CR flux, $j_h$, is anisotropic due
to our position at 8.5 kpc from the center of the Galaxy. The
anisotropy can be obtained from eq.(\ref{j-halo}),
\[
j_h(\theta) \propto \int dx \, n(r(x,\theta)).
\]
Fig.1a shows the anisotropy $j_h(0)/j_h(\pi)$ as a function of the
core radius for the trial distributions 
\begin{equation}
n(r)  \propto  {1\over (r_c^2 +r^2)}
\label{n=1/r^2}
\end{equation}
and  
\begin{equation}
n(r) \propto  {1\over \sqrt{(r_c^2 +r^2)} (R_h+r)^2},
\label{n=1/r}
\end{equation}
where $R_h$ is the halo size.  First of these distributions describes
isothermal halo model \cite{halo1} while the second one is more
realistic distribution of ref.~\cite{halo2}.  We have arbitrarily
regularized it at $r=0$ by introducing the core size $r_c$.  For
homogeneous distribution $n(x)=\mbox{const}\times\theta(R_h-r)$ the
anisotropy is minimum and constitutes about 20\%.  Fig.1b shows
corresponding angular dependencies of $j_h(\theta)$ at $r_c=5$ kpc. As
can be seen from the picture, the anisotropy of the galactic
contribution is at least $\sim 20$\% and can be much larger if $n(x)$
is concentrated around the galactic center. Also, it should be noted
that the anisotropy depends exclusively on $n(x)$ and does not depend
on energy since cosmic rays with energy $E\sim E_{\scriptsize\rm GZK}$
are deflected by the Galactic magnetic field by $\sim 3^{\circ}$ at 
most~\cite{deflection}.

In
order to see the significance of galactic part, it is
necessary to compare the galactic component $j_h$ of the total flux with the
isotropic extragalactic contribution $j_{ext}$. By making
use of eq.(\ref{n/n_h}), one obtains
\begin{equation}
{j_{\scriptsize\rm ext} \over j_h} =  
\alpha {R_{\scriptsize\rm ext} \over R_h} { \bar n \over \bar
n_h} \sim \alpha {R_{\scriptsize\rm ext} \over R_h} \times 10^{-5} ,  
\label{jext/jh}
\end{equation}
where $R_h\sim 100$~kpc is the size of the Galactic halo and $\alpha$
is the constant of purely geometrical origin,
\begin{equation}
\alpha = { 3\int_{r<R_h} d^3x n(x)\over R_h^2
\int_{r<R_h} {d^3x \over x^2} n(x)} . 
\label{alpha}
\end{equation}
Here $r(x,\theta)=(x^2+r_0^2-2xr_0\cos\theta)^{1/2}$ is the distance
between current point and the Galactic center while $r_0=8.5$ kpc
is the distance to the Galactic center.
The numerical value of $\alpha$ is $\alpha\simeq 0.15$ and
$\alpha\simeq 0.5$ for distributions (\ref{n=1/r^2}) and
(\ref{n=1/r}), respectively, with no strong dependence on $r_c$ in the
range $r_c = 2 - 10$ kpc, while for homogeneous distribution 
$\alpha$ is close to 1.

From eq.(\ref{jext/jh}) one finds
\begin{eqnarray}
{j_{\scriptsize\rm ext} \over j_h} &\sim& \alpha
\makebox[1in]{}\mbox{for $E<E_{\scriptsize\rm GZK}$}, \nonumber \\ 
{j_{\scriptsize\rm ext} \over j_h}
&\sim& 10^{-2} \times \alpha
\makebox[1.2cm]{}\mbox{for $E>E_{\scriptsize\rm GZK}$}.  
\label{jext/jh-estimate}
\end{eqnarray}
Therefore, at $E<E_{\scriptsize\rm GZK}$ the Galactic and
extragalactic contributions can be comparable (although the Galactic
one is probably somewhat larger), while at $E>E_{\scriptsize\rm GZK}$
the extragalactic part is suppressed by a factor $\sim 10^{-2}$. In
either case a substantial fraction of the observed UHE CR should come
from the halo of our Galaxy. In this respect our conclusions agree
with that of ref.\cite{X-Ber}.

Since at energies above the GZK cutoff the extragalactic contribution
is negligible, non-observation of the anisotropy at the level of $\sim
20$\% would rule out the CDM-related mechanisms of UHE CR. The
observation of the Galactic anisotropy would allow to reconstruct the
density profile $n(x)$ and, possibly, the distribution of CDM in the
Galactic halo. 

At energies below the GZK cutoff, the anisotropy is smaller due to the
relative enhancement of the isotropic extragalactic part. The latter
should have narrow peaks in the direction of nearby galaxies and clusters.
The contribution of
such a peak, $\delta j_{\scriptsize\rm ext}$, equals
\[
{\delta j_{\scriptsize\rm ext}\over j_h} =  \alpha
{R_h^2\over 3 R^2} {M\over M_G}, 
\]
where $R$ is the distance to the astronomical object, $M$ is its mass,
and $M_G$ is the mass of our Galaxy including halo. For instance,
contributions from Andromeda Nebula and Virgo Cluster are comparable
and close to $10^{-2}\times \alpha$, in agreement with
eq.(\ref{jext/jh-estimate}) and ref.\cite{X-Ber}.

Since anisotropy does not depend on energy and can be measured at
$E>E_{\scriptsize\rm GZK}$, it is possible, in principle, to determine
the relative magnitude of the extragalactic contribution. Provided the
CDM-related mechanisms are dominant at $E\lesssim E_{\scriptsize\rm
GZK}$ and the coefficient $\alpha$ is known, the ratio
$j_h/j_{\scriptsize\rm ext}$ could give, in view of eqs.(\ref{n/n_h})
and (\ref{jext/jh}), an important information about the distribution
of matter in the Universe.

Current data are not enough to draw definite conclusions about the
angular distribution of highest energy cosmic rays both because
of very limited statistics and the absence of data in the South
hemisphere where the Galactic center is situated. However, since the
anisotropy predicted by the CDM-related mechanisms is large, it will
be either observed or excluded already in the next generation of
experiments \cite{experiments}. Among these the Pierre Auger Project
has the best potential due to large number of expected events
(600--1000 events with $E>10^{20}$ eV in 10 years) and the ability to
see both hemispheres.

In conclusion, it is worth noting that there are mechanisms
in which smaller but still observable galactic anisotropy is expected. 
As an example, consider a model based on annihilation of
high energy neutrinos on massive relic neutrinos~\cite{Waxman}.
Relic neutrinos with mass $m_{\nu}\sim 1$~eV are non-relativistic at
present and may be expected to cluster in galactic halos similar to
CDM. 
From the Pauli exclusion principle, the maximum number density of
neutrinos
in the Galactic halo scales with neutrino mass as $(m_{\nu})^3$. At 
$m_{\nu}\sim 10$~eV corresponding flux from the halo of our Galaxy is
10 times bigger than the extragalactic one, which may lead to
observable anisotropy.

\paragraph{Acknowledgments.}
The authors would like to thank D.~S.~Gorbunov, V.~A.~Kuz\-min,
V.~A.~Rubakov, M.V.~Sazhin and D.V.~Semikoz for helpful
discussions. The work of P.T. is supported in part by Award
No. RP1-187 of the U.S. Civilian Research \& Development Foundation
for the Independent States of the Former Soviet Union (CRDF), and by
Russian Foundation for Basic Research, grants 96-02-17804a.The work of
S.D. is supported in part by Russian Foundation for Basic Research
grant 96-02-17449a, by the INTAS grant 96-0457 within the research
program of the International Center for Fundamental Physics in Moscow
and by ISSEP fellowship.

\def\ijmp#1#2#3{{\it Int. Jour. Mod. Phys. }{\bf #1~}(19#2)~#3}
\def\pl#1#2#3{{\it Phys. Lett. }{\bf B#1~}(19#2)~#3}
\def\zp#1#2#3{{\it Z. Phys. }{\bf C#1~}(19#2)~#3}
\def\prl#1#2#3{{\it Phys. Rev. Lett. }{\bf #1~}(19#2)~#3}
\def\rmp#1#2#3{{\it Rev. Mod. Phys. }{\bf #1~}(19#2)~#3}
\def\prep#1#2#3{{\it Phys. Rep. }{\bf #1~}(19#2)~#3}
\def\pr#1#2#3{{\it Phys. Rev. }{\bf D#1~}(19#2)~#3}
\def\np#1#2#3{{\it Nucl. Phys. }{\bf B#1~}(19#2)~#3}
\def\mpl#1#2#3{{\it Mod. Phys. Lett. }{\bf #1~}(19#2)~#3}
\def\arnps#1#2#3{{\it Annu. Rev. Nucl. Part. Sci. }{\bf #1~}(19#2)~#3}
\def\sjnp#1#2#3{{\it Sov. J. Nucl. Phys. }{\bf #1~}(19#2)~#3}
\def\jetp#1#2#3{{\it JETP Lett. }{\bf #1~}(19#2)~#3}
\def\app#1#2#3{{\it Acta Phys. Polon. }{\bf #1~}(19#2)~#3}
\def\rnc#1#2#3{{\it Riv. Nuovo Cim. }{\bf #1~}(19#2)~#3}
\def\ap#1#2#3{{\it Ann. Phys. }{\bf #1~}(19#2)~#3}
\def\ptp#1#2#3{{\it Prog. Theor. Phys. }{\bf #1~}(19#2)~#3}
\def\spu#1#2#3{{\it Sov. Phys. Usp.}{\bf #1~}(19#2)~#3}

\end{document}